\documentclass[aps,pre,preprint,showpacs,amsmath,amssymb]{revtex4}
\usepackage{graphicx} 

\begin{document}

\title{Stochastic nonlinear differential equation generating $1/f$ noise}

\author{B. Kaulakys}
\email{kaulakys@itpa.lt}
\author{J. Ruseckas}
\affiliation{Institute of Theoretical Physics and Astronomy, Vilnius
University,\\ A. Go\v{s}tauto 12, LT-01108 Vilnius, Lithuania}

\date{\today{}}

\begin{abstract}
Starting from the simple point process model of $1/f$ noise we
derive a stochastic nonlinear differential equation for the signal exhibiting
$1/f$ noise in any desirably wide range of frequency. A stochastic differential
equation (the general Langevin equation with a multiplicative noise) that gives
$1/f$ noise is derived for the first time. The solution of the equation exhibits
the power-law distribution. The process with $1/f$ noise is demonstrated by the
numerical solution of the derived equation with the appropriate restriction of
the diffusion of the signal in some finite interval.
\end{abstract}

\maketitle

The power spectra of a large variety of systems ranging widely from astrophysics
and technology to sociology and psychology at low frequencies have $1/f$
behavior, i.e., the power density $S(f)$ is inversely proportional to the
frequency $f$
\cite{Press78,Hooge81,Dutta81,Weissman88,Zhigalskii97,Gilden95,Thurner97,Scher02,Milotti02}.
$1/f$ noise, also known as flicker noise, is intermediate between white noise
(no correlation in time, $S(f)\sim1/f^0$) and the Brownian motion (no
correlation between increments, $S(f)\sim1/f^2$). Simple procedures of
integration or differentiation of such fluctuating signals do not yield the
signal exhibiting $1/f$ noise. Most of the $1/f$ noise models are specialized
or complicated. This makes the problem of omnipresent $1/f$ noise one of the
oldest puzzles in the contemporary physics. In contrast to the Brownian motion generated 
by the linear stochastic equation, simple systems of differential,
even linear stochastic equations generating signals with $1/f$ noise are not known.

The purpose of this paper is the derivation of a nonlinear stochastic differential
equation (generalized Langevin equation for the signal) generating signal with
$1/f$ noise. The stochastic differential equation is obtained from the point
process model of $1/f$ noise, analyzed in
Refs.~\cite{Kaulakys981,Kaulakys982,Kaulakys991,Kaulakys992,Kaulakys001,Kaulakys002,
Kaulakys003}. Such a method enables one to obtain various stochastic
differential equations, starting from different point processes and generating
stochastic signals with different slopes of the power density. Analysis of the concrete physical models and application of the derived nonlinear stochastic equation for modeling of the specific observable processes are out of the scope of this paper. 

We start from the point process model recently proposed and analyzed in
Refs.~\cite{Kaulakys981,Kaulakys982,Kaulakys991,Kaulakys992,Kaulakys001,Kaulakys002,Kaulakys003}.
The signal in the model consists of pulses or series of events,
\begin{equation}
I(t)=a\sum_k\delta(t-t_k).\label{eq:point}
\end{equation}
Here $\delta(t)$ is the Dirac delta function, $\{t_k\}$ is a set of the
occurrence times at which the particles or pulses cross the section of
observation and $a$ is a contribution to the signal of one pulse or particle.
The power spectral density of the point process (\ref{eq:point}) may be
expressed as
\cite{Kaulakys981,Kaulakys982,Kaulakys991,Kaulakys992,Kaulakys001,Kaulakys002,Kaulakys003}
\begin{equation}
S(f)=\lim_{T\rightarrow\infty}\left\langle\frac{2a^2}{T}\sum_{k,q}e^{i2\pi
 f\Delta(k;q)}\right\rangle ,
\end{equation}
where $T$ is the observation time and
\begin{equation}
\Delta(k;q)\equiv t_{k+q}-t_k=\sum_{l=k}^{k+q-1}\tau_l
\end{equation}
is the difference between the pulses occurrence times $t_{k+q}$ and $t_k$. Here
the brackets $\langle\dots\rangle$ denote the averaging over the realizations of
the process and $\tau_k=t_{k+1}-t_k$ is the interevent time. In the model
\cite{Kaulakys981,Kaulakys982,Kaulakys991,Kaulakys992,Kaulakys001,Kaulakys002,Kaulakys003}
the interevent time of the signal stochastically diffuses about some
average value and the process has been described by an autoregressive iteration
with a very small relaxation. Here we will consider the stochastic point process
described by the recurrent equations
\begin{eqnarray}
t_{k+1} & = & t_k+\tau_k,\label{eq:tk}\\
\tau_{k+1} & = &\tau_k+\sigma\varepsilon_k \label{eq:tauk}
\end{eqnarray}
with the appropriate boundary conditions, restricting the diffusion of $\tau_k$
in the finite interval $[\tau_{\mathrm{min}},\tau_{\mathrm{max}}]$. In
Eq.~(\ref{eq:tauk}) $\varepsilon_k$ are normally distributed uncorrelated random
variables with a zero expectation and unit variance, i.e., a white noise, and
$\sigma$ is a standard deviation of the white noise.

\begin{figure}
\includegraphics[width=0.90\textwidth]{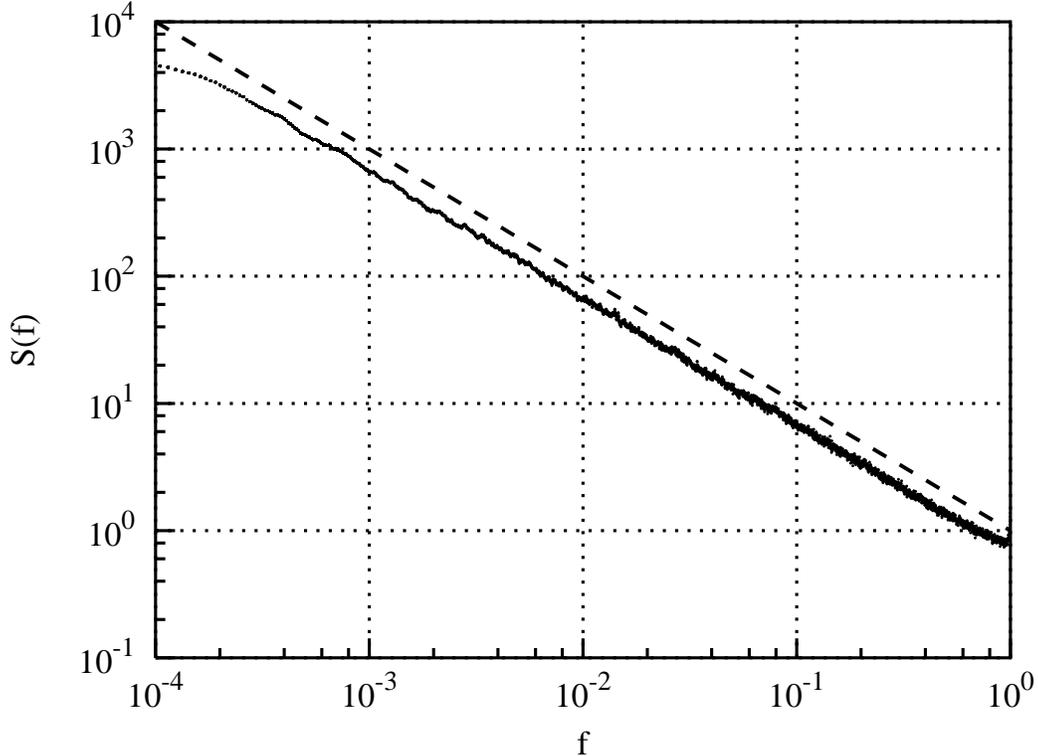}
\caption{Power spectral density of the point process, described by the equations
(\ref{eq:tk}) and (\ref{eq:tauk}). Parameters used are $\sigma=0.01$,
$\tau_{\mathrm{min}}=10^{-5}$, and $\tau_{\mathrm{max}}=1$. The dashed line
represents the spectral density, calculated according to equation $S(f)=1/f$.}
\label{fig:point-spectr}
\end{figure}

The signal (\ref{eq:point}) generated according to Eqs.~(\ref{eq:tk}) and
(\ref{eq:tauk}), depending on the parameter $\sigma$ and the interval
$[\tau_{\mathrm{min}},\tau_{\mathrm{max}}]$, exhibits $1/f$ noise in any
desirably wide range of frequency. According to the general theory
\cite{Kaulakys981,Kaulakys982,Kaulakys991,Kaulakys992,Kaulakys001,Kaulakys002,Kaulakys003}
the power spectral density of such point process for $f\lesssim\tau_{
\mathrm{max}}^{-1}$ and $\tau_{\mathrm{min}}\rightarrow0$ may be estimated as
\begin{equation}
S(f)\sim\frac{a^2}{\tau_{\mathrm{max}}^2}\frac{1}{f}\,.\label{eq:spectrdiskr}
\end{equation}
The spectrum obtained from the numerical solution of Eqs.~(\ref{eq:tk}) and
(\ref{eq:tauk}) with reflective boundary conditions at $\tau_{\mathrm{min}}$ and
$\tau_{\mathrm{max}}$ is shown in Fig.~\ref{fig:point-spectr}. We see that the
considered point process gives $1/f$ noise in a wide range of frequencies.

It is the purpose of this paper to derive a stochastic differential equation
for the signal, the solution of which exhibits $1/f$ noise. For this purpose we
rewrite Eq.~(\ref{eq:tauk}) as a differential Ito stochastic equation
interpreting $k$ as a continuous variable, i.e.,
\begin{equation}
\frac{d\tau_k}{dk}=\sigma\xi(k).\label{eq:dtaudk}
\end{equation}
Here $\xi(k)$ is a Gaussian white noise satisfying the standard condition
\begin{equation}
\langle\xi(k)\xi(k')\rangle=\delta(k-k').
\end{equation}
 Then we rewrite Eq.~(\ref{eq:dtaudk}) using the occurrence time. Transition
from the occurrence number $k$ to the actual time $t$ according to the relation
$dt=\tau_kdk$ yields equation
\begin{equation}
\frac{d\tau}{dt}=\frac{\sigma}{\sqrt{\tau}}\xi(t).\label{eq:taudiff}
\end{equation}

The signal averaged over the time interval $\tau_k$ according to
Eq.~(\ref{eq:point}) is $x=a/\tau_k$. The standard \cite{Gardiner85}
transformation of the variable from $\tau$ to $x=a/\tau$ in
Eq.~(\ref{eq:taudiff}) results in the stochastic differential Ito equation
\begin{equation}
\frac{dx}{dt}=\frac{\sigma^2}{a^3}x^4+\frac{\sigma}{a^{3/2}}x^{
5/2}\xi(t).\label{eq:xdiff}
\end{equation}
Equation (\ref{eq:xdiff}) can be rewritten in the form that does not contain any
parameters. Introducing the scaled time
\begin{equation}
t_s=\frac{\sigma^2}{a^3}t
\end{equation}
we obtain from Eq.~(\ref{eq:xdiff}) an equation
\begin{equation}
\frac{dx}{dt_s}=x^4+x^{5/2}\xi(t_s).\label{eq:xscaled}
\end{equation}

The steady state solution of the stationary Fokker-Planck equation with the
appropriate reflective boundary conditions and a zero flow obtained from
Eq.~(\ref{eq:xscaled}) according to the standard method \cite{Gardiner85} is of
the power-law form,
\begin{equation}
P(x)=\frac{C}{x^3},\label{eq:power}
\end{equation}
where $C$ has to be defined from the normalization.

The power-law distribution of the signals is the phenomenon observable in a
large variety of processes from earthquakes to the financial time series
\cite{Thurner97,Scher02,Bak02,Gabaix03}. Therefore, our model of $1/f$ noise is
complementary to the models based on the superposition of signals with a wide-range
distribution of the relaxation times resulting into the Gaussian process
\cite{McWhorter57}.

Because of the divergence of the power-law distribution and requirement of the
stationarity of the process, the stochastic equation (\ref{eq:xscaled}) should
be analyzed together with the appropriate restrictions of the diffusion in some
finite interval $x_{\mathrm{min}}\lessapprox x\lessapprox x_{\mathrm{max}}$.
Such restrictions may be introduced as some additional conditions to the
iterative solution of the stochastic differential equation. The similar
restrictions, however, may be fulfilled by introducing some additional terms into
Eq.~(\ref{eq:xscaled}), corresponding to the restriction of the diffusion in
some {}``potential well''. According to the general theory \cite{Gardiner85} the
exponentially restricted diffusion with the distribution density
\begin{equation}
P(x)\sim\frac{1}{x^3}\exp\left\{-\left(\frac{x_{\mathrm{min}}}{x}\right)^n
-\left(\frac{x}{x_{\mathrm{max}}}\right)^n\right\}\label{eq:distr}
\end{equation}
generates the stochastic differential equation
\begin{equation}
\frac{dx}{dt_s}=\frac{n}{2}\left(\frac{x_{\mathrm{min}}^n}{x^{n-4}}-\frac{x^{n
+4}}{x_{\mathrm{max}}^n}\right)+x^4+x^{5/2}\xi(t_s).\label{eq:restricted}
\end{equation}
Here $n$ is some parameter.

Equations (\ref{eq:xscaled}) and (\ref{eq:restricted}) are the main result of
this letter. Since the point process (\ref{eq:tk}) and (\ref{eq:tauk}) gives the
signal with $1/f$ noise, the signal obtained from Eqs.~(\ref{eq:xscaled}) and
(\ref{eq:restricted}) also should give $1/f$ noise in some frequency interval.
When $x_{\mathrm{max}}\rightarrow\infty$, from Eq.~(\ref{eq:spectrdiskr}) we can
estimate the power spectral density as
\begin{equation}
S(f)\sim x_{\mathrm{min}}^2\frac{1}{f}\,.
\end{equation}
 Such a conclusion is confirmed by the numerical solution of
Eq.~(\ref{eq:restricted}).

\begin{figure}[htbp]
\includegraphics[width=0.90\textwidth]{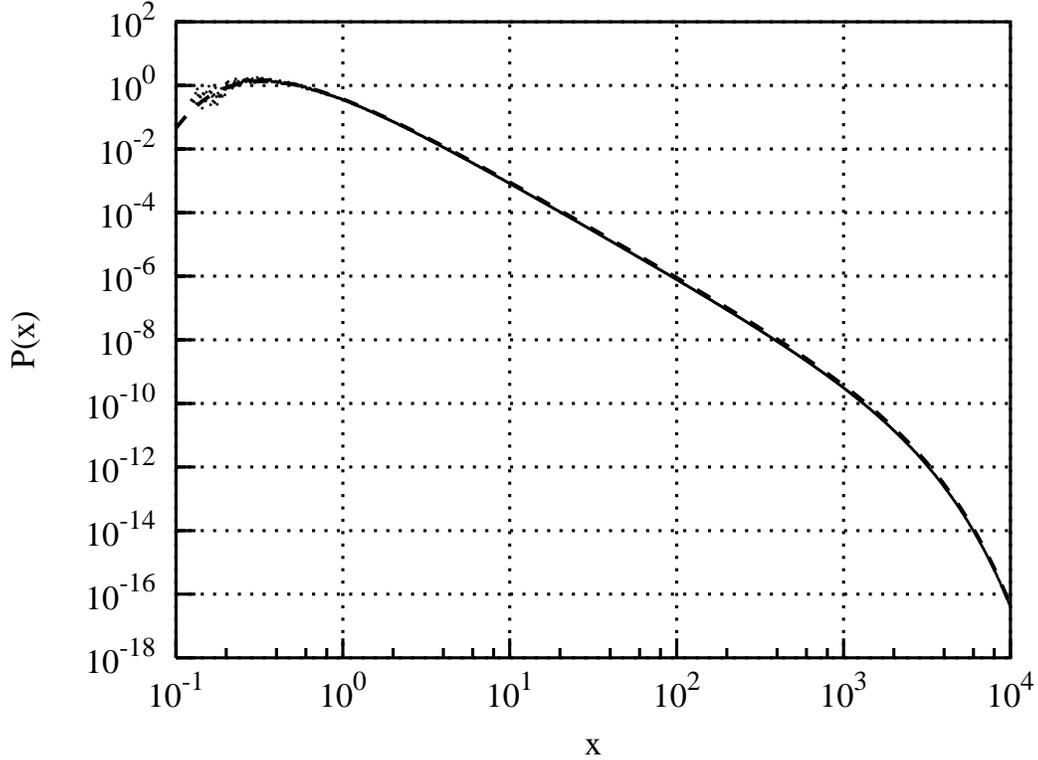}
\caption{Distribution density $P(x)$ obtained from the numerical solution of
Eq.~(\ref{eq:proport}). The dashed line represents the distribution density
calculated from Eq.~(\ref{eq:distr}). Parameters used are $x_{\mathrm{min}}=1$,
$x_{\mathrm{max}}=10^3$, $n=1$, and $\kappa=0.1$.}
\label{fig:distrib}
\end{figure}

\begin{figure}[htbp]
\includegraphics[width=0.90\textwidth]{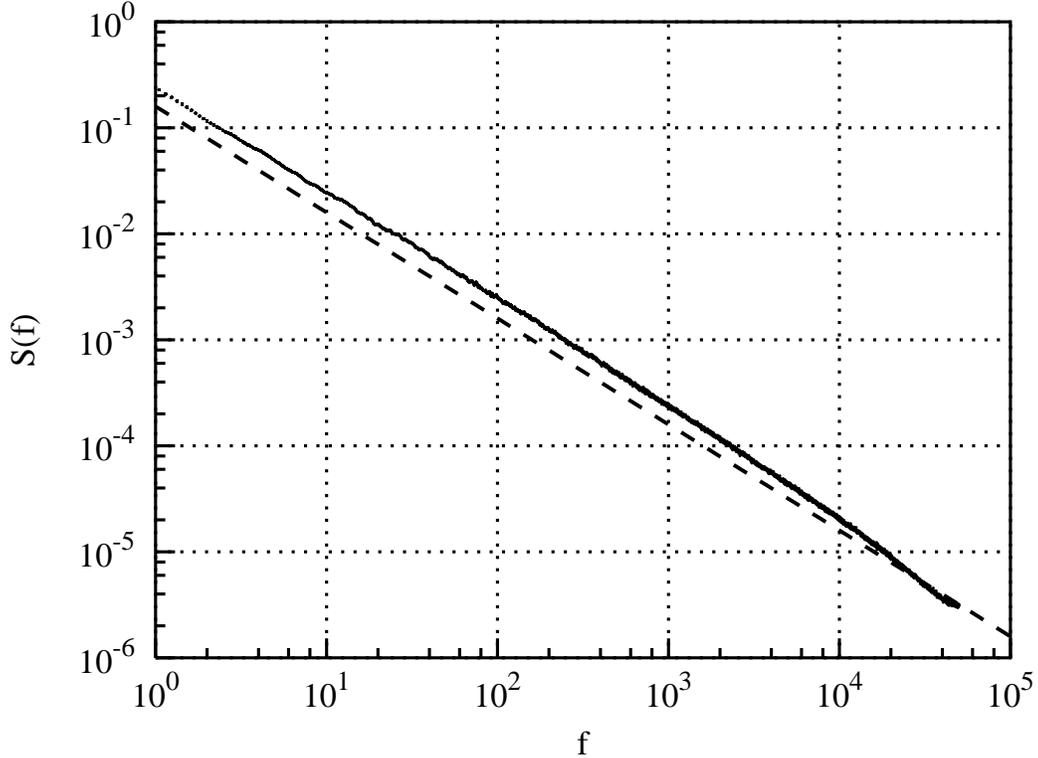}
\caption{Power spectral density of the signal, obtained from
Eq.~(\ref{eq:proport}). The dashed line represents the spectral density,
calculated according to equation $S(f)=1/(2\pi f)$. The parameters used are the
same as in Fig.~\ref{fig:distrib}.}
\label{fig:spectr-diff}
\end{figure}

We solve Eqs.~(\ref{eq:xscaled}) and (\ref{eq:restricted}) using the method of
discretization. When the variable step of integration is $\Delta t_s=h_i$, the 
differential equation (\ref{eq:restricted}) transforms to the difference equation 
\begin{equation}
x_{i+1}=x_i+\frac{n}{2}\left(\frac{x_{\mathrm{min}}^n}{x_i^{n-4}}-\frac{x_i^{n
+4}}{x_{\mathrm{max}}^n}\right)h_i+x_i^4h_i+x_i^{5/2}\sqrt{
h_i}\varepsilon_i.
\label{eq:discrete}
\end{equation}
We can solve Eq.~(\ref{eq:discrete}) numerically with the constant step,
$h_i=const$, when $t_{i+1}=t_i+h$. However, one of the most effective method of
solutions of Eq.~(\ref{eq:discrete}) is when the change of the variable $x_i$ in
one step is proportional to the value of the variable. We take the integration
steps $h_i$ from the equation $x_i^{5/2}\sqrt{h_i}=\kappa x_i$ with $\kappa\ll1$
being a small parameter. As a result we have the system of the equations
\begin{eqnarray}
x_{i+1} & = & x_i+\kappa^2x_i\left[1+\frac{n}{2}\left(\frac{x_{\mathrm{min}}^n}{
x_i^n}-\frac{x_i^n}{x_{\mathrm{max}}^n}\right)\right]+\kappa
 x_i\varepsilon_i,\label{eq:proport}\\
t_{i+1} & = & t_i+\frac{\kappa^2}{x_i^3}.\nonumber
\end{eqnarray}

The distribution density $P(x)$ of the variable $x$, obtained using
Eq.~(\ref{eq:proport}) is shown in Fig.~\ref{fig:distrib}. We see that our
method of solution gives good agreement with the power-law distribution
(\ref{eq:power}) in the interval $x_{\mathrm{min}}\lesssim x\lesssim
x_{\mathrm{max}}$.

The power spectral density $S(f)$ calculated
according to Eqs.~(\ref{eq:proport}) is shown in Fig.~\ref{fig:spectr-diff}.
Figure \ref{fig:spectr-diff} shows that equation (\ref{eq:restricted}) indeed
gives a signal exhibiting $1/f$ noise in a wide frequency interval.

In summary, for the first time we derived a stochastic differential equation for
the signal exhibiting $1/f$ noise in any desirably wide range of frequency. The
distribution density of the signal is of the inverse cubic power-law. The
numerical analysis of the obtained equation shows that the signal indeed
exhibits $1/f$ noise and power-law distribution. 

The research described in this publication was supported in part by the Lithuanian State and Studies Foundation. 


\end{document}